\documentclass[prd,twocolumn,showpacs,amsmath,amssymb]{revtex4}

\usepackage{graphicx}
\usepackage{dcolumn}
\usepackage{bm}
\usepackage{psfrag}
\usepackage{subfigure}

\newcommand{\be}{\begin{eqnarray}}
\newcommand{\ee}{\end{eqnarray}}
\newcommand{\bn}{\begin{eqnarray*}}
\newcommand{\en}{\end{eqnarray*}}

\newcommand{\nn}{\nonumber \\}
\newcommand{\nl}{\\}

\renewcommand{\vec}[1]{\mbox{\boldmath$#1$}}

\newcommand{\gslash}[1]{\mbox{\slash{\hspace{-2mm}}$#1$}}

\renewcommand{\th}{\ensuremath{\theta}}

\newcommand{\ph}{\ensuremath{\phi}}
\newcommand{\vph}{\ensuremath{\varphi}}
\newcommand{\al}{\ensuremath{\alpha}}
\newcommand{\bt}{\ensuremath{\beta}}
\newcommand{\sg}{\ensuremath{\sigma}}
\newcommand{\gm}{\ensuremath{\gamma}}

\newcommand{\gfive}{\ensuremath{\gm^5}}

\newcommand{\pvec}{\ensuremath{\vec{p}}}
\newcommand{\Pvec}{\ensuremath{\vec{P}}}
\newcommand{\nvec}{\ensuremath{\vec{n}}}
\newcommand{\kvec}{\ensuremath{\vec{k}}}

\newcommand{\Rvec}{\ensuremath{\vec{R}}}

\newcommand{\alvec}{\ensuremath{\vec{\al}}}
\newcommand{\sgvec}{\ensuremath{\vec{\sg}}}

\newcommand{\Vvec}{\ensuremath{\vec{V}}}

\newcommand{\Wvec}{\ensuremath{\vec{W}}}
\newcommand{\Ovec}{\ensuremath{\vec{\Omega}}}

\newcommand{\nabvec}{\ensuremath{\vec{\nabla}}}

\newcommand{\hb}{\ensuremath{\hbar}}
\newcommand{\ihb}{\ensuremath{i \hbar}}
\newcommand{\pt}[1]{\ensuremath{{\partial \over \partial #1}}}

\newcommand{\lt}{\ensuremath{\left}}
\newcommand{\rt}{\ensuremath{\right}}
\newcommand{\dotprod}[2]{\ensuremath{\lt(\vec{#1} \cdot \vec{#2}\rt)}}

\begin{document}

\pagenumbering{arabic}

\title{The implications of noninertial motion on covariant quantum spin}%

\author{Dinesh Singh}
\email{singhd@uregina.ca}
\affiliation{%
Department of Physics, University of Regina \\
Regina, Saskatchewan, S4S 0A2, Canada
}%
\author{Nader Mobed}
\email{nader.mobed@uregina.ca}
\affiliation{%
Department of Physics, University of Regina \\
Regina, Saskatchewan, S4S 0A2, Canada
}%
\date{\today}

\begin{abstract}

It is shown that the Pauli-Lubanski spin vector defined in terms of curvilinear co-ordinates
does not satisfy Lorentz invariance for spin-1/2 particles in noninertial motion
along a curved trajectory.
The possibility of detecting this violation in muon decay experiments is explored, where
the noninertial contribution to the decay rate becomes large for muon beams with large momenta
and trajectories with radius of curvature approaching the muon's Compton wavelength scale.
A new spacelike spin vector is derived from the Pauli-Lubanski vector that
satisfies Lorentz invariance for both inertial and noninertial motion.
In addition, this spin vector suggests a generalization for the classification of
spin-1/2 particles, and has interesting properties that are applicable for both
massive and massless particles.


\end{abstract}


\maketitle

\section{Introduction}

Quantum mechanics and general relativity are regarded as the two pillars of modern theoretical physics,
but despite many attempts in finding a satisfactory theory of quantum gravity, this goal has proved elusive.
Understanding how the two theories can fit together harmoniously is a serious intellectual challenge.
Approaches such as string theory \cite{Polchinski,Green}, loop quantum gravity \cite{Ashtekar,Rovelli},
twistor theory \cite{Penrose}, causal set theory \cite{Sorkin}, and others
each assume distinct mathematical and philosophical foundations, and necessarily possess a high degree of
mathematical sophistication.
Whether any one of these approaches can ultimately lead to the definitive theory of quantum gravity is purely
speculative at present because there is no observational evidence to support or disregard them.
While there is yet no experimental or observational justification to tamper with either quantum mechanics
or general relativity, there are suggestions that a modification of one or both theories may be required to incorporate
Planck-scale physics.
Whether such modifications are justifiable or not is also debatable.
Because both quantum mechanics and general relativity are very successful in their respective domains, it is a
challenge to propose a modification without compelling evidence to support the change.

Any sort of observation involving gravity and matter at the microscopic level is notoriously difficult to
perceive, let alone perform, which makes the advancement of quantum gravity research extremely challenging outside of
purely theoretical considerations.
From investigating quantum field theory in a classical space-time background, the topic of black hole entropy and the
prediction of Hawking radiation \cite{Hawking} from black holes is widely cited as an indicator of an effect that a successful quantum
gravity theory must reproduce.
Similarly, the analogous prediction of thermal radiation observed by an accelerated observer in flat space-time,
now known as the Unruh effect \cite{Unruh}, suggests a type of confirmation that supports this viewpoint, via the principle
of equivalence.
However, a closer inspection of certain outstanding issues \cite{Helfer} involving Hawking radiation may indicate that
inherent assumptions about the nature of space-time at the quantum level need to be very clearly identified
and addressed.
If it turns out that a subtle change is required on the assumptions for how particles behave in curved backgrounds,
it may force one to question the reliability of quantum field theory in curved space-time and the consequences that
follow, such as Hawking radiation.


One avenue where the interface between quantum mechanics and general relativity may be explored involves the intrinsic
nature of particles in a curved space-time background.
From standard quantum field theory in flat space-time, the classification of subatomic particles according to their
mass and spin angular momentum via the Poincar\'{e} symmetry group \cite{Ryder} is a well known and accepted procedure.
Yet it must be emphasized that the Poincar\'{e} group is applicable only for particles in strictly inertial motion.
In reality, the notion of a truly inertial reference frame is merely an idealization,
since general relativity demonstrates that any mass-energy source generates space-time curvature,
which produces a gravitational field.
At best, the closest approximation to an inertial frame in a gravitational field is the so-called
{\it freely falling frame}, in which a particle propagates along a geodesic and feels no external forces while
on this worldline.
Even here, however, it is impossible to completely neutralize the gravitational field effects for any particle
that is not strictly pointlike, since an extended object can sense tidal forces acting upon it due to space-time
curvature, as determined by neighbouring geodesics along the object's centre-of-mass worldline.
As well, it is certainly possible to describe noninertial motion in Minkowski space-time, where
the deviations from force-free motion are treated classically as effects due to ``fictitious'' forces.
These facts indicate that the Poincar\'{e} group is insufficient for the description of intrinsic properties of
subatomic particles while in noninertial motion.

Although it is currently difficult to envision a suitable generalization of the Poincar\'{e} group to operate on a
general curved space-time background, there exists one type of generalization called the {\it de~Sitter group}
\cite{Gursey} for the special case of conformally flat space-times in four dimensions.
Because the de~Sitter space-time can be embedded in a flat five-dimensional space-time described by an
overall radius of curvature $R$, the momentum operators are elements of a ten-parameter set of rotation generators.
The Casimir invariants of the de~Sitter group are then generalizations of the squared mass and spin
involving $R$, which then reduce to their corresponding expressions for the Poincar\'{e} group as
$R \rightarrow \infty$.
However, it also appears that the de~Sitter group suffers from the same limitations as the Poincar\'{e} group,
in that its properties rely heavily on inertial motion of the particle to obtain the associated Casimir invariants.
This is because, in all known formulations of such space-time symmetry groups, cartesian co-ordinates are automatically
assumed for the co-ordinates and derivative operators to compose the group generators.
Because physical properties should be independent of the choice of co-ordinate system, it seems reasonable to
re-examine the Poincar\'{e} and de~Sitter groups assuming general curvilinear co-ordinates.

The purpose of this paper is to investigate the properties of one aspect of the Poincar\'{e} group in
curvilinear co-ordinates, namely the Pauli-Lubanski four-vector which describes spin-1/2 particle spin in covariant form.
If a particle's trajectory appears as a curved path in three-dimensional space, it can be mapped by a
non-cartesian co-ordinate system which best reflects the symmetry of the particle's motion.
By now describing the Pauli-Lubanski vector with respect to curvilinear co-ordinates, it may be
possible to see whether a breakdown occurs in describing the particle's associated Casimir invariant,
as determined by some inertial laboratory frame defined at the origin.
It must be clearly understood that this paper is not an attempt to establish new theory, but rather to
examine the phenomenological consequences of exploring the limitations of the Poincar\'{e} group when applied to
noninertial motion.
If there is a significant {\em observational} deviation that appears from exploring this perspective,
then it seems reasonable to propose a modification to accommodate the difference and explore the consequences
that follow.

The paper begins in Sec.~II with a brief description on the hypothesis of locality employed for the description of
particle motion in a noninertial reference frame.
This is followed by Sec.~III, which describes the Pauli-Lubanski vector for a spin-1/2 particle
generalized in terms of curvilinear co-ordinates, with interesting consequences when the intended Casimir invariants are evaluated.
An application of the main results is presented in Sec.~IV, which consider the effects of noninertial motion on muon decay
in a circular storage ring.
It is then shown in Sec.~V that a new form of covariant spin operator can be constructed from the Pauli-Lubanski vector
which satisfies the Casimir invariant conditions for both inertial and noninertial motion, with interesting properties to follow.
This leads to a conclusion in Sec.~VI, which explores possible future developments of this investigation.

\section{The Hypothesis of Locality for Noninertial Motion}
\begin{figure*}
\psfrag{x}[cl][][2.5][0]{\tiny ${\rm x}$}
\psfrag{y}[cr][][2.5][0]{\tiny ${\rm y}$}
\psfrag{th}[cc][][2.5][0]{\tiny $\ph$}
\psfrag{x1}[bl][][2.5][0]{\tiny $\hat{x}$}
\psfrag{y1}[tb][][2.5][0]{\tiny $\hat{y}$}
\psfrag{r1}[bl][][2.5][0]{\tiny $\hat{r}$}
\psfrag{th1}[bl][][2.5][0]{\tiny $\hat{\ph}$}
\begin{minipage}[t]{0.3 \textwidth}
\centering
\subfigure[\hspace{0.2cm} cartesian]{
\label{fig:cartesian}
\rotatebox{0}{\includegraphics[width = 5cm, height = 5cm, scale = 1]{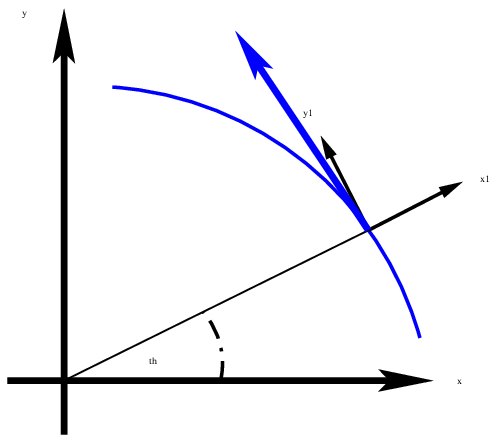}}}
\end{minipage}%
\hspace{2.0cm}
\begin{minipage}[t]{0.3 \textwidth}
\centering
\subfigure[\hspace{0.2cm} polar]{
\label{fig:polar}
\rotatebox{0}{\includegraphics[width = 5cm, height = 5cm, scale = 1]{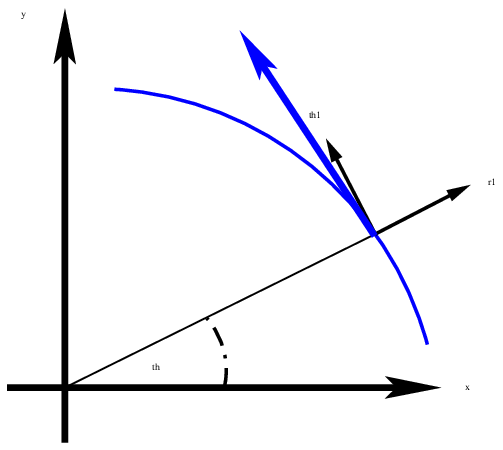}}}
\end{minipage}
\caption{\label{fig:co-ordinate} Comoving co-ordinate frames for a quantum particle moving along a curved trajectory.
For muon decay, it is possible to identify the beam's instantaneous location with either local cartesian co-ordinates $(\hat{x}, \hat{y}, \hat{z})$
or cylindrical polar co-ordinates $(\hat{r}, \hat{\vph}, \hat{z})$ with respect to the laboratory frame $\rm (x, y, z)$,
as shown in Figures~\ref{fig:cartesian} and \ref{fig:polar}.
However, the fundamental issue is whether one local frame is more appropriate than the other for accurately mapping the
muon beam's trajectory, with implications for describing potential non-inertial effects.}
\end{figure*}
A central assumption used in describing noninertial motion is the
{\it hypothesis of locality} \cite{Mashhoon}, in which an accelerated observer can be equated with an
instantaneously comoving inertial observer at a given moment of proper time.
This is described mathematically by defining a family of velocity vectors tangent to each point on the observer's
worldline, which smoothly correspond one-to-one with their respective comoving inertial frames.
Embedded with this assumption is an intrinsic length and time scale associated with the accelerated observer.
The hypothesis of locality is certainly reasonable for an accelerated classical particle,
since its instantaneous position and velocity occupy the same state as that of comoving inertial particle.
As well, in the limit as the particle becomes pointlike, this hypothesis is also well-defined.
However, the situation becomes more complicated when applied to a quantum mechanical system because
the quantum particle has wavelike properties and requires a {\it region} of space-time to define its location properly.
It is possible to surmise that the hypothesis of locality may become violated when considering accelerated quantum systems possessing
long wavelengths.
Conversely, it is possible to retain this hypothesis if the time scale associated with the measurement of
a quantum particle is much shorter than the accelerated observer's intrinsic time scale.

\section{General Formalism}
Assuming $c = 1$ units throughout, a flat metric in terms of curvilinear co-ordinates and $-2$ signature is described by
$\vec{g} = \eta_{\hat{\mu}\hat{\nu}} \, \vec{e}^{\hat{\mu}} \otimes \vec{e}^{\hat{\nu}} \,$,
given a set of orthonormal tetrads
$\lt\{\vec{e}_{\hat{\mu}} \rt\}$ and basis one-forms $\lt\{\vec{e}^{\hat{\mu}} \rt\}$
labelled by caratted indices to define a local Minkowski frame satisfying
$\lt\langle \vec{e}^{\hat{\mu}} , \vec{e}_{\hat{\nu}} \rt\rangle =
\delta^{\hat{\mu}}{}_{\hat{\nu}} \,$.
A general metric tensor in curved space-time and its Minkowski counterpart are related
by vierbein sets $\lt\{ e^{\hat{\alpha}}{}_\mu \rt\} \,$,
$\lt\{ e^\mu{}_{\hat{\alpha}} \rt\}$ satisfying
$\vec{e}^{\hat{\alpha}} = e^{\hat{\alpha}}{}_\beta \, \vec{e}^\beta$ and
$\vec{e}_{\hat{\alpha}} = e^\beta{}_{\hat{\alpha}} \, \vec{e}_\beta \,$, such that
$e^{\hat{\alpha}}{}_\mu \, e^\mu{}_{\hat{\beta}} =
\delta^{\hat{\alpha}}{}_{\hat{\beta}} \, $,
$
e^{\mu}{}_{\hat{\alpha}} \, e^{\hat{\alpha}}{}_{\nu} =
\delta^\mu{}_\nu \, $,
%
and
$g_{\mu \nu} = \eta_{\hat{\alpha}\hat{\beta}} \,
e^{\hat{\alpha}}{}_\mu \, e^{\hat{\beta}}{}_\nu$.
%
The caratted indices are raised and lowered by the Minkowski metric $\eta_{\hat{\al}\hat{\bt}}$, while
the uncaratted indices are manipulated with the general metric $g_{\mu \nu}$.

Given the covariant Dirac equation
\be
\lt[i \gm^\mu(x) \lt[\partial_\mu + i \, \Gamma_\mu(x) \rt] - m/\hb\rt]\psi(x) & = & 0 \,
\label{Dirac-eq}
\ee
with the spin connection $\Gamma_\mu(x)$, the associated momentum operator is
$\Pvec^\mu = i \hbar \, g^{\mu \nu}\lt[\partial_\nu + i \, \Gamma_\nu(x)\rt] =
e^\mu{}_{\hat{\mu}} \, \Pvec^{\hat{\mu}}$, where $\Pvec^{\hat{\mu}} = \pvec^{\hat{\mu}} + \Ovec^{\hat{\mu}}$
is described in terms of the operator
$\pvec^{\hat{\mu}} = \ihb \, \nabvec^{\hat{\mu}}$ in curvilinear co-ordinates \cite{Hassani} and
the spin connection contribution $\Ovec^{\hat{\mu}} \,$.
The $\pvec$-momentum components are explicitly described by
%
\be
\pvec^{\hat{0}} & = & \ihb \, \pt{t} \, ,
\qquad
\pvec^{\hat{\jmath}}
\ = \ - {\ihb \over \lambda^j(u)} \, \pt{u^{\hat{\jmath}}} \, ,
\label{momentum-curve}
\ee
%
where $\lambda^j(u)$ are the scale functions corresponding to the co-ordinate system that best reflects
the symmetries of the particle's motion in space.
Consider now the Pauli-Lubanski four-vector $\Wvec^\mu = e^\mu{}_{\hat{\mu}} \, \Wvec^{\hat{\mu}}$
in curved space-time as a vierbein projection onto a local tangent space, such that \cite{Itzykson}
\be
\Wvec^{\hat{\mu}} & = & -{1 \over 4} \, \varepsilon^{\hat{\mu}}{}_{\hat{\al}\hat{\bt}\hat{\gm}} \,
\sg^{\hat{\al}\hat{\bt}} \, \Pvec^{\hat{\gm}} \, ,
\label{Wvec-defn}
\ee
where $\Wvec^{\hat{\mu}}$ is the Pauli-Lubanski vector in Minkowski space-time,
$\varepsilon^{\hat{\mu}\hat{\al}\hat{\bt}\hat{\gm}}$ is the Levi-Civita alternating symbol with
$\varepsilon^{0123}~\equiv~1$, and
$\sg^{\hat{\al}\hat{\bt}} = \lt(i/2\rt)[\gm^{\hat{\al}}, \gm^{\hat{\bt}}]$
are the spin matrices.
By introducing a decomposed representation of $\sg^{\hat{\al}\hat{\bt}}$ in the form
\be
\sg^{\hat{\al}\hat{\bt}} & = &
i \lt(\delta^{\hat{\al}}{}_0 \, \delta^{\hat{\bt}}{}_{\hat{\jmath}} -
\delta^{\hat{\bt}}{}_0 \, \delta^{\hat{\al}}{}_{\hat{\jmath}} \rt) \alvec^{\hat{\jmath}}
- \varepsilon^{0\hat{\al}\hat{\bt}}{}_{\hat{m}} \, \sgvec^{\hat{m}} \, ,
\label{sg-defn}
\ee
where $\alvec^{\hat{\jmath}} = \bt \, \gamma^{\hat{\jmath}}$,
$\sgvec^{\hat{\jmath}} = {1 \over 2} \, \epsilon^j{}_{kl} \, \sg^{\hat{k}\hat{l}}$, and
$\epsilon^{ijk}$ is the three-dimensional Levi-Civita alternating symbol with $\epsilon^{123}~\equiv~1$,
it follows naturally that
\be
\Wvec^{\hat{\mu}} & = & {1 \over 2} \lt[-\eta^{\hat{\mu} \hat{0}} \, \sgvec_{\hat{m}} \, \Pvec^{\hat{m}} \rt.
\nn
& &{} + \lt. \eta^{\hat{\mu} \hat{m}} \lt[\sgvec_{\hat{m}} \, \Pvec^{\hat{0}}
+ i \, \epsilon_{mjk} \, \alvec^{\hat{\jmath}} \, \Pvec^{\hat{k}} \rt] \rt] \, .
\label{Wvec}
\ee

Because it is generally true that $\lt(i/\hb\rt)\lt[\pvec^{\hat{\imath}}, \pvec^{\hat{\jmath}}\rt] \equiv
N^{\hat{\imath}\hat{\jmath}}~\neq~0 \,$ in curvilinear co-ordinates, there exists a Hermitian three-vector
$\Rvec$ called the {\em noninertial dipole operator} \cite{Singh1,Singh2}, whose more general form in
terms of $\Ovec$ is now
\be
\Rvec^{\hat{k}} & = & {i \over 2\hbar} \, \epsilon^{ijk} \, [\Pvec_{\hat{\imath}}, \Pvec_{\hat{\jmath}}]
\nn
& = & \epsilon^{kmn} \lt[\lt(\nabvec_{\hat{m}} \, \ln \lambda^n\rt) \pvec_{\hat{n}} - \nabvec_{\hat{m}} \Ovec_{\hat{n}}\rt]\, .
\label{Rvec}
\ee
For example, if the locally flat tangent space-time is described by spherical co-ordinates, then for
\be
\Pvec^{\hat{r}} \ = \ - \ihb \lt(\pt{r} + {1 \over r}\rt) \, , \quad
\Pvec^{\hat{\th}} & = & - {\ihb \over r} \lt(\pt{\th} + {1 \over 2} \, \cot \th\rt) \, ,
\nn
\Pvec^{\hat{\ph}} & = & - {\ihb \over r \, \sin \th} \, \pt{\ph} \, ,
\label{momentum1}
\ee
it follows that
\begin{subequations}
\be
\Rvec^{\hat{r}} & = & {i \over \hb}\lt[\Pvec^{\hat{\th}}, \Pvec^{\hat{\ph}}\rt] \ = \
- {\cot \th \over r} \, \pvec^{\hat{\ph}} \, ,
\label{Rvec1}
\nl
\Rvec^{\hat{\th}} & = & {i \over \hb}\lt[\Pvec^{\hat{\ph}}, \Pvec^{\hat{r}}\rt] \ = \
{1 \over r} \, \pvec^{\hat{\ph}} \, ,
\label{Rvec2}
\nl
\Rvec^{\hat{\ph}} & = & {i \over \hb}\lt[\Pvec^{\hat{r}}, \Pvec^{\hat{\th}}\rt] \ = \
- {1 \over r} \, \Pvec^{\hat{\th}} \, ,
\label{Rvec3}
\ee
\end{subequations}
where $r$ is interpreted as the local radius of curvature associated with the trajectory
of the particle.
As $r~\rightarrow~\infty$, it is self-evident that $\Rvec \rightarrow \vec{0}$
for fixed momentum.
If the $\Pvec^{\hat{\jmath}}$ are defined by cartesian co-ordinates to represent rectilinear motion, then $\Rvec$
vanishes identically, justifying its label to describe noninertial effects due to rotation.

For a spin-1/2 particle in a rotating frame, it can be shown that while
$\Pvec^\mu \, \Pvec_\mu = m^2$ remains a Lorentz scalar, the corresponding scalar operator
associated with particle spin is now {\it frame-dependent}, in the form
\be
\Wvec^\mu \, \Wvec_\mu & = & \Wvec_\mu \, \Wvec^\mu
\nn
& = &
-{1 \over 2} \lt({1 \over 2} + 1\rt)m^2  + {\hb \over 2} \dotprod{\sg}{R} \, .
\label{W^2}
\ee
In addition, the orthogonality condition between $\Wvec$ and $\Pvec$ is no longer
satisfied for noninertial motion due to rotation, since
\be
\Wvec^\mu \, \Pvec_\mu & = & -\Pvec^\mu \, \Wvec_\mu \ = \ -{\hb \over 2} \dotprod{\al}{R} \, .
\label{W.p}
\ee
%
%
It may seem contradictory to now regard the Casimir invariant $\Wvec^\mu \, \Wvec_\mu$ as a frame-dependent quantity
when it is constructed as a Lorentz scalar.
However, when this operator is applied to an {\it eigenstate}, the associated eigenvalue is $-{1 \over 2} \lt({1 \over 2} + 1\rt)m^2$,
and any contribution due to $\Rvec$ does not appear.
As well, $\Wvec^\mu \, \Pvec_\mu = 0$ when applied to an eigenstate.
For both (\ref{W^2}) and (\ref{W.p}), the apparent violation of Lorentz symmetry and the
orthogonality condition is due to a spin interaction with $\Rvec$, which suggests that noninertial motion induces a
spin precession about the particle's worldline.

\begin{figure}
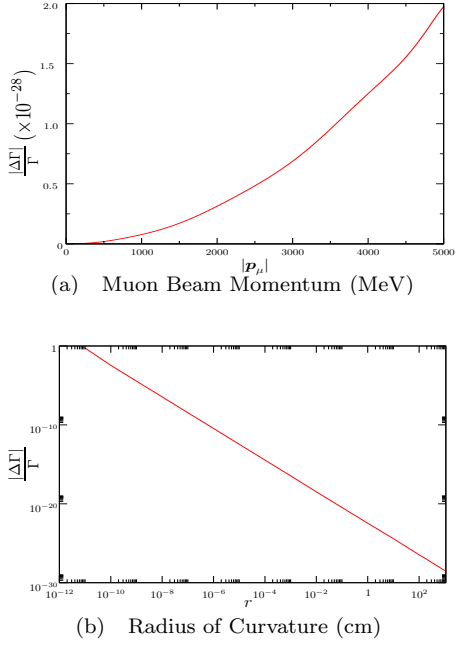

\psfrag{dG/G}[cc][][4][0]{${|\Delta \Gamma| \over \Gamma} \, {(\times 10^{-28})}$}
\psfrag{dG/G1}[cc][][4][0]{${|\Delta \Gamma| \over \Gamma}$}
\psfrag{r (cm)}[tr][][2][0]{\Large $r$}
\psfrag{Muon Beam Momentum (MeV)}[tc][][2][0]{\Large $|\pvec_\mu|$}
\psfrag{0}[tc][][2][0]{$\tiny 0$}
\psfrag{1000}[tc][][2][0]{$1000$}
\psfrag{2000}[tc][][2][0]{$2000$}
\psfrag{3000}[tc][][2][0]{$3000$}
\psfrag{4000}[tc][][2][0]{$4000$}
\psfrag{5000}[tc][][2][0]{$5000$}
\psfrag{100}[tc][][2][0]{$10^2$}
\psfrag{1}[tc][][2][0]{$1$}
\psfrag{0.01}[tc][][2][0]{$10^{-2}$}
\psfrag{0.0001}[tc][][2][0]{$10^{-4}$}
\psfrag{1e-06}[tc][][2][0]{$10^{-6}$}
\psfrag{1e-08}[tc][][2][0]{$10^{-8}$}
\psfrag{1e-10}[tc][][2][0]{$10^{-10}$}
\psfrag{1e-12}[tc][][2][0]{$10^{-12}$}
\psfrag{1e-20}[tc][][2][0]{$10^{-20}$}
\psfrag{1e-30}[tc][][2][0]{$10^{-30}$}
\psfrag{5e-29}[tc][][2][0]{$0.5$}
\psfrag{1e-28}[tc][][2][0]{$1.0$}
\psfrag{1.5e-28}[tc][][2][0]{$1.5$}
\psfrag{2e-28}[tc][][2][0]{$2.0$}
\begin{minipage}[t]{0.3 \textwidth}
\centering
\subfigure[\hspace{0.2cm} Muon Beam Momentum (MeV)]{
\label{fig:NR1a}
\rotatebox{0}{\includegraphics[width = 5.8cm, height = 3.5cm, scale = 1]{2a}}}
\end{minipage}%
\vspace{0.4cm}
\begin{minipage}[t]{0.3 \textwidth}
\centering
\subfigure[\hspace{0.2cm} Radius of Curvature (cm)]{
\label{fig:NR1b}
\rotatebox{0}{\includegraphics[width = 5.7cm, height = 3.5cm, scale = 1]{2b}}}
\end{minipage}
\caption{\label{fig:NR1} Relative contribution of $\Rvec$ on the decay rate for $\mu^-~\rightarrow~e^-~+~\bar{\nu}_e~+~\nu_\mu$,
where $r = 711.2$~cm for Fig.~\ref{fig:NR1a}, while $|\pvec_\mu| = 3.0$~GeV for Fig.~\ref{fig:NR1b}.
It is evident 
that $|\Delta \Gamma|/\Gamma$ becomes large as $r \rightarrow 10^{-12}$ cm,
the length scale of the muon's Compton wavelength.}
\end{figure}

\section{Applications to Muon Decay}

One of the assumptions in quantum field theory often taken for granted is the notion that every elementary
interaction within a Feynman diagram occurs at a {\it mathematical point}, where it is assumed that momentum
conservation is precisely satisfied at each vertex.
While it appears that this hypothesis is valid for interactions on a flat space-time background,
this idealization has to be examined more carefully when applied to a curved background or in situations
involving noninertial motion.
While elementary particles are reasonably assumed to be pointlike objects, quantum fields exist as a continuous distribution of matter,
and since a gravitational field is described classically as tidal forces within a local neighbourhood about
some point on the manifold, there are conceptual ambiguities involved in how best to represent the interactions of
quantum matter in a gravitational field \cite{Birrell}.
Unlike any collision involving classical objects where the properties of the incident and final products of the reaction are completely
determined, the Heisenberg uncertainty principle implies that an ambiguity exists at the precise moments when a scattering
or decay event occurs.

One example of how $\Rvec$ may appear within an experimental context is via muon decay
inside a circular storage ring.
If an event like muon decay is only well-defined up to some identifiable interaction region, then it is reasonable
to suspect that $\Rvec$ is a legitimate operator which leads to a meaningful prediction of physical effects.
For inertial motion, it is well-known that the computed rest-frame decay rate \cite{Scadron,Kaku}
for the reaction $\mu^- \rightarrow e^- + \bar{\nu}_e + \nu_\mu$ is
\be
\Gamma_0 & \approx & {G_{\rm F}^2 \, m_\mu^5 \over 192 \, \pi^3} \ \approx \ 2.965 \times 10^{-16} \ {\rm MeV},
\label{G0}
\ee
where $G_{\rm F} = 1.16639 \times 10^{-11}$~MeV$^{-2}$ \cite{Kaku} is the Fermi constant for weak interactions.
If the muon beam follows a circular trajectory and the decay products travel inertially off the orbit,
then the squared modulus of the interaction matrix element takes the form
\be
|{\cal M}|^2 & = & |{\cal M}|_0^2 + |{\cal M}|_{\rm NI}^2 \ = \
{G_{\rm F}^2 \over 2} \, L_{\hat{\mu} \hat{\nu}} \, M^{\hat{\mu} \hat{\nu}},
\label{M^2}
\ee
where $|{\cal M}|_0^2$ is the known contribution due to inertial motion \cite{Kaku}, and $|{\cal M}|_{\rm NI}^2$
is the correction due to $\Rvec$.
The tensors corresponding to this reaction are then
\begin{subequations}
\be
\hspace{-0.8cm}
L^{\hat{\mu} \hat{\nu}} & = & {\rm Tr} \lt[\gslash{\pvec}_{\nu_\mu} \, \gm^{\hat{\mu}} \lt(\gslash{\Pvec}_\mu
+ m_\mu \, \gfive \, \gslash{\nvec}_\mu\rt)\gm^{\hat{\nu}} \lt(1 - \gfive\rt)\rt] , \,
\label{L_uv}
\nl
\hspace{-0.8cm}
M^{\hat{\mu} \hat{\nu}} & = & {\rm Tr} \lt[\lt(\gslash{\pvec}_e + m_e \, \gfive \, \gslash{\nvec}_e\rt)
\gm^{\hat{\mu}} \, \gslash{\pvec}_{\nu_e} \, \gm^{\hat{\nu}} \lt(1 - \gfive\rt)\rt] ,
\label{M_uv}
\ee
\end{subequations}
where $\nvec_\mu$ is the muon's spin polarization vector.
Prior to averaging over the spin states and using (\ref{W.p}), it can be shown that \cite{Itzykson}
\be
\gfive \, \gslash{\nvec}_\mu & = &
- {2 \over m_\mu^2} \lt[\lt(\nvec \cdot \Wvec\rt)\gslash{\Pvec} + \gslash{\nvec} \lt(\Pvec \cdot \Wvec\rt)\rt]_\mu \, .
\label{slash_nvec}
\ee
By substituting (\ref{slash_nvec}) into (\ref{L_uv}), it follows that
\be
\lefteqn{
|{\cal M}|_{\rm NI}^2 \ = \ {20 \, G_{\rm F}^2 \, \hbar^2 \, E_\mu \over m_\mu^2 \, \lt| \Pvec_\mu \rt|}
\lt[\lt[\lt(1 - {E_e \over |\pvec_e|}\rt) \lt(\pvec_e \cdot \pvec_{\nu_\mu} \rt) \rt. \rt. }
\nn
&&{} + \lt. \lt. {m_e^2 \over |\pvec_e|} \, E_{\nu_\mu} \rt]
\pvec_{\nu_e}^{\hat{\jmath}}
- {3 \over 5}\lt(1 - {E_e \over |\pvec_e|}\rt)\lt(\pvec_{\nu_\mu} \cdot \pvec_{\nu_e}\rt)\pvec_e^{\hat{\jmath}}\rt]
\nn
&&{} \times \epsilon_{jkl} \lt(\nabvec^{\hat{k}} \Rvec^{\hat{l}}\rt) \, ,
\label{M^2-noninertial}
\ee
which leads to the noninertial contribution $\Delta \Gamma$ for the total decay rate $\Gamma = \Gamma_0 + \Delta \Gamma$
in the laboratory frame.
It is interesting to note the prediction of an induced coupling of the neutrinos due to rotational motion,
an effect not present in $|{\cal M}|_0^2$.

Adopting cylindrical co-ordinates for the muon motion in the laboratory frame, where
$\Pvec_\mu^{\hat{\jmath}} = \lt(\Pvec_\mu^{\hat{z}}, \Pvec_\mu^{\hat{r}}, \Pvec_\mu^{\hat{\varphi}}\rt)$,
the only nonzero component of the noninertial dipole operator is
$\Rvec^{\hat{1}} = -{1 \over r} \, \pvec_\mu^{\hat{\varphi}}$ \cite{Singh2}.
Figure~\ref{fig:co-ordinate} provides a comparative description of the muon beam's instantaneous location in terms of
either a local cartesian frame in Figure~\ref{fig:cartesian} or the adopted cylindrical co-ordinate frame in Figure~\ref{fig:polar}.
When evaluated in (\ref{M^2-noninertial}), it follows that $\Delta \Gamma$ is negative-valued, suggesting that
the muon's lifetime becomes enhanced when moving along a circular trajectory.
The degree of enhancement is strongly dependent on the choice of $r$, as shown in Figure~\ref{fig:NR1}, where
Figs.~\ref{fig:NR1a} and \ref{fig:NR1b} predict the relative contribution of $\Delta \Gamma$ to the overall decay rate
for varying beam momenta and radius of curvature, respectively.
For a facility like the Brookhaven National Laboratory which performs the muon $g-2$ experiment \cite{Farley},
the magnitude of $\Rvec$ is negligibly small, where for a muon beam momentum of
$|\pvec_\mu| = 3.094$~GeV and storage ring radius of $r = 711.2$~cm, $|\Rvec| = 5.760 \times 10^{-55} \ll 1$.

By this example, it seems that $\Rvec$ is irrelevant for current particle experiments.
However, as $r \sim 10^{-8} - 10^{-12}$~cm, the length scales associated with atomic and nuclear dimensions \cite{Krane},
$|\Delta \Gamma|/\Gamma$ rapidly approaches unity.
In particular, the total decay rate becomes {\em negative} for $r \lesssim 10^{-11}$~cm, when the length scale
approaches the muon's Compton wavelength of $1.18 \times 10^{-12}$ cm \cite{Griffiths}.
It is difficult to make a sensible physical interpretation of such a decay rate at the Compton wavelength scale,
and it seems reasonable to suspect that some known physical effect, such as bremmstrahlung radiation,
will manifest itself to avoid this scenario.
At the same time, this may be a sign of new physics emerging at the interface between quantum mechanics and general relativity,
consistent with reported evidence claiming, for example, that space-time curvature can induce modifications
to classical electromagnetic fields at the Compton wavelength scale \cite{Rosquist}.

It is worthwhile to consider at this stage whether the external magnetic field required to curve the muon's trajectory
will influence the particle's decay rate.
For example, the external magnetic field in a storage ring will, in principle, modify the decay rate via the density of
final states.
This effect has been examined in detail for the case of neutron beta decay in a strong magnetic field \cite{Fassio-Canuto},
where it is shown that a field strength in excess of $10^{12}$~Gauss is required to significantly modify the neutron's
decay rate.
In particular, the electron's final state density reflects the decay rate corrections in terms of an overall scale factor,
which can be taken into account in any realistic measurement.
Applying this formalism to muon decay, it is found that the relative correction to the overall scale factor
is negligibly small for a typical magnetic field strength of $10^4$~Gauss required to bend the particle beam.

Concerning a more direct examination based on a leading order calculation \cite{van-Holten}, it has been shown
that the relative correction to the muon's lifetime is on the order of $10^{-14}$ times the
magnetic field strength measured in Tesla.
Based on this assumption, an external magnetic field of $10^{11}$ Tesla or $10^{15}$ Gauss is required to
achieve a radius of curvature approaching the muon's Compton wavelength scale of $10^{-12}$ cm.
It is interesting to note that this prior calculation leads to a correction of the muon's lifetime with the same
order of magnitude as shown in this paper, while not accounting for the possible existence of $\Rvec$ described by (\ref{Rvec}).
Such a finding suggests that a much more detailed analysis may be required to better understand the implications
for either a breakdown of currently applied formalism or the emergence of new physical phenomena at this length scale.




\section{An Equivalent Expression for the Pauli-Lubanski Vector for Noninertial Motion}


Having shown that the squared magnitude of the Pauli-Lubanski vector in curvilinear co-ordinates is no longer a Lorentz
invariant for noninertial motion along a curved trajectory, is it possible to construct a new vector that satisfies this property?
It so happens that it is possible to obtain a spin vector that preserves Lorentz invariance for general particle motion.
Consider $\Vvec^\mu = e^\mu{}_{\hat{\mu}} \, \Vvec^{\hat{\mu}}$, where
%
\be
\Vvec^{\hat{\mu}} & \equiv & A \, \Wvec^{\hat{\mu}} + B \, \eta^{\hat{\mu}\hat{0}} \, \sgvec_{\hat{m}} \, \Pvec^{\hat{m}}
\nn
& &{} + \eta^{\hat{\mu}\hat{m}} \lt[C \, \sgvec_{\hat{m}} \, \Pvec^{\hat{0}}
+ i \, D \, \epsilon_{mjk} \, \alvec^{\hat{\jmath}} \, \Pvec^{\hat{k}} \rt] \, ,
\label{Vvec-defn}
\ee
with $A, B, C, D$ as algebraic constants to be evaluated.
If $C$ and $D$ are chosen such that
$C = -{A \over 2} \pm {i \, \sqrt{3} \over 6} \lt(A - 2 \, B\rt)$ and
$D = -{A \over 2} \pm {i \over 2} \lt(A - 2 \, B\rt) \, $,
then it follows that
\be
\Vvec^{\hat{\mu}} & = & {\kappa \over 2} \lt[i \, \eta^{\hat{\mu}\hat{0}} \, \sgvec_{\hat{m}} \, \Pvec^{\hat{m}} \rt.
\nn
& &{} \pm \lt. \eta^{\hat{\mu}\hat{m}} \lt[{1 \over \sqrt{3}} \, \sgvec_{\hat{m}} \, \Pvec^{\hat{0}}
+ i \, \epsilon_{mjk} \, \alvec^{\hat{\jmath}} \, \Pvec^{\hat{k}}  \rt]  \rt]
\nn
& = & {\kappa \over 2} \lt[-2 \, i \, \Wvec^{\hat{\mu}} \pm \eta^{\hat{\mu}\hat{m}}
\lt[\lt({1 \over \sqrt{3}}  \mp i\rt) \sgvec_{\hat{m}} \, \Pvec^{\hat{0}} \rt. \rt.
\nn
& &{} + \lt. \lt. \lt(1 \pm i \rt)i \, \epsilon_{mjk} \, \alvec^{\hat{\jmath}} \, \Pvec^{\hat{k}} \rt] \rt] \, ,
\label{Vvec}
\ee
and
\be
\Vvec^\mu \, \Vvec_\mu & = & - {1 \over 4} \, \kappa^2 \, m^2 \,
\label{V^2}
\ee
for real-valued $\kappa \equiv i \lt(A - 2 \, B\rt)$.
If $\kappa = \pm \sqrt{3}$, then
$\Vvec^\mu \, \Vvec_\mu = \lt(-3/4\rt)m^2 = \lt[-{1 \over 2}\lt({1 \over 2} + 1\rt)\rt]m^2$, resembling the Lorentz
invariance condition described by $\Wvec^\mu \, \Wvec_\mu$, but now applicable for all forms of motion.

In particular, if $A \equiv 1$ in (\ref{Vvec-defn}) for $\kappa = \pm \sqrt{3}$, then it follows that
$B = {1 \over 2} \pm {i \sqrt{3} \over 2}$, $C = 1$, and $D = {\sqrt{3} \over 2} - {1 \over 2}$.
By expressing $B$ in complex exponential form, it is shown that
\be
\Vvec^{\hat{\mu}} & = & \Wvec^{\hat{\mu}} + e^{\pm i \xi} \, \eta^{\hat{\mu}\hat{0}} \, \sgvec_{\hat{m}} \, \Pvec^{\hat{m}}
\nn
& &{} + \eta^{\hat{\mu}\hat{m}} \lt[\sgvec_{\hat{m}} \, \Pvec^{\hat{0}}
+ i \, \sqrt{2} \sin \lt(\xi - {\pi \over 4}\rt) \, \epsilon_{mjk} \, \alvec^{\hat{\jmath}} \, \Pvec^{\hat{k}} \rt] \, ,
\nn
\label{Vvec-A=1}
\ee
where $\xi = {\pi \over 3}$.  As well, by letting $B = |B|\lt(\cos \xi + i \sin \xi \rt)$ for arbitrary and complex-valued $\kappa$,
it follows that
\be
\xi & = & \sin^{-1} \lt({{\rm Re}(\kappa) \over 2 |B|}\rt) \ = \ \cos^{-1} \lt({1 - {\rm Im}(\kappa) \over 2 |B|}\rt).
\label{xi=}
\ee

A preliminary study of $\Vvec$, defined in terms of (\ref{Vvec}), reveals some interesting properties worth considering.
The first is that, in combination with (\ref{W.p}), the projection of the new spin operator along $\Pvec$ generates
the helicity operator up to the phase $\xi = {\pi \over 3}$, where
%
%
\be
\lt(\Vvec^\mu \mp \kappa \, \Wvec^\mu\rt)\Pvec_\mu & = &
\Pvec^\mu\lt(\Vvec_\mu \mp \kappa \, \Wvec_\mu\rt)
\nn
& = & \mp \, {\kappa \, E \over 2}\lt({1 \over \sqrt{3}} \pm i\rt) \dotprod{\sg}{P}
\nn
& = & \mp \, {\kappa \, E \over \sqrt{3}} \, e^{\pm i \xi} \, \dotprod{\sg}{P} ,
\label{(V+W).p}
\ee
%
%
where $E$ is the energy eigenvalue of $\Pvec^{\hat{0}}$.
The second property is that, for a polarization vector defined as
\be
\nvec^{\hat{\mu}} & = & {E \over m} \lt({\lt| \Pvec \rt| \over E}, {1 \over \lt| \Pvec \rt|} \, \Pvec^{\hat{\jmath}}\rt)
\label{nvec}
\ee
subject to $\nvec^\mu \, \Pvec_\mu = 0$, it can be shown that a new vector
\be
\kvec^{\hat{\mu}} & = & {2E \over m^2} \lt({1 \over \lt| \Pvec \rt|} \, \Pvec^{\hat{\mu}} - {m \over E} \, \nvec^{\hat{\mu}} \rt)
\label{kvec}
\ee
can project out the helicity operator from both $\Vvec$ and $\Wvec$.
That is,
\be
\kvec^\mu \, \Wvec_\mu \ = \
{i \over \kappa} \, \kvec^\mu \, \Vvec_\mu & = & {\dotprod{\sg}{P} \over \lt| \Pvec \rt|} \, ,
\label{orthog1}
\nl
\kvec^\mu \lt(\Vvec_\mu + i \, \kappa \, \Wvec_\mu \rt) & = & 0 \, ,
\label{orthog2}
\ee
where $\kvec^{\hat{\mu}} = \lt(2 / \lt| \Pvec \rt|\rt) \, \delta^{\hat{\mu}}{}_{\hat{0}}$.
This expression shows that the helicity operator is rotated away from the particle's direction of propagation
due to noninertial motion along a curved trajectory.
Moreover, for massless particles, a null polarization vector can be defined as
$\nvec^{\hat{\mu}} = \lt(1, {1 \over \lt| \Pvec \rt|} \, \Pvec^{\hat{\jmath}} \rt)$ orthogonal to $\Pvec$, such that
$\kvec^\mu \, \Wvec_\mu = \kvec^\mu \, \Vvec_\mu = 0$, where
\be
\kvec^{\hat{\mu}} & = & \Pvec^{\hat{\mu}} - \lt| \Pvec \rt| \, \nvec^{\hat{\mu}}.
\label{kvec-null}
\ee
This indicates that $\Vvec^\mu = \lambda \, \kvec^\mu$ for massless spin-1/2 particles, an expression now generalized
to account for both inertial and noninertial motion, where $\lambda = \pm 1/2$ is the associated helicity eigenvalue.

In addition, (\ref{(V+W).p}) is true for both massive and {\em massless}
$\lt(\Pvec^\mu \, \Pvec_\mu = 0\rt)$ spin-1/2 particles.
To show this explicitly, consider the special case of a massless particle propagating in the 3-direction
$\Pvec^{\hat{\mu}} = \lt(E, 0, 0, E\rt)$.
Then
\begin{subequations}
\be
\Vvec^{\hat{0}} & = & -{i \, \kappa \, E \over 2} \, \sgvec^{\hat{3}} \ = \ \mp \, i \sqrt{3} \, \Vvec^{\hat{3}} \, ,
\label{Vvec0}
\nl
\Vvec^{\hat{1}} & = & \pm \, {\kappa \, E \over 2} \lt[{1 \over \sqrt{3}} \, \sgvec^{\hat{1}} + i \, \alvec^{\hat{2}} \rt] \, ,
\label{Vvec1}
\nl
\Vvec^{\hat{2}} & = & \pm \, {\kappa \, E \over 2} \lt[{1 \over \sqrt{3}} \, \sgvec^{\hat{2}} - i \, \alvec^{\hat{1}} \rt] \, ,
\label{Vvec2}
\nl
\Vvec^{\hat{3}} & = & \pm \, {\kappa \, E \over 2} \, {1 \over \sqrt{3}} \, \sgvec^{\hat{3}}
\ = \ \pm \, {i \over \sqrt{3}} \, \Vvec^{\hat{0}} \, ,
\label{Vvec3}
\ee
\end{subequations}
from which it follows naturally that
\be
\Vvec^\mu \, \Pvec_\mu & = & \Vvec^{\hat{0}} \, \Pvec_{\hat{0}} + \Vvec^{\hat{3}} \, \Pvec_{\hat{3}}
\nn
& = & \mp \, {\kappa \, E \over 2}\lt({1 \over \sqrt{3}} \pm i\rt) \lt(\sgvec^{\hat{3}} \, \Pvec^{\hat{3}}\rt) \, ,
\label{V.p}
\ee
to agree with (\ref{(V+W).p}) for inertial motion.

The third interesting feature comes from examining the commutation properties of (\ref{Vvec0})-(\ref{Vvec3}).
After setting $\kappa = \pm \sqrt{3}$ and re-defining the generators such that $\vec{\cal V} \equiv \Vvec/E$,
it becomes evident that
\begin{subequations}
\be
{[\vec{\cal V}^{\hat{1}}, \vec{\cal V}^{\hat{2}}]} & = & - 6 \, i \, \vec{\cal V}^{\hat{3}} \, ,
\label{SU(2)-a}
\nl
{[\vec{\cal V}^{\hat{2}}, \vec{\cal V}^{\hat{3}}]} & = & i \, \vec{\cal V}^{\hat{1}} \, ,
\label{SU(2)-b}
\nl
{[\vec{\cal V}^{\hat{3}}, \vec{\cal V}^{\hat{1}}]} & = & i \, \vec{\cal V}^{\hat{2}} \, ,
\label{SU(2)-c}
\ee
\end{subequations}
%
%
%
which generates the Lie algebra for a broken $SU(2)$ symmetry.
This suggested property of massless spin-1/2 particles, while perhaps more physically intuitive, is quite
different from what is known from Wigner's little group procedure \cite{Ryder,Scadron}, since the latter indicates
that massless particle rotation is described by the Euclidean group $E(2)$.
The reasons for this discrepancy are unclear at present.

\section{Conclusion}

It has been shown in this paper that the Pauli-Lubanski covariant spin vector $\Wvec^\mu$ in curvilinear co-ordinates is not a
Casimir operator, where the breakdown comes from noninertial effects due to rotation.
As well, it is shown that there exists a new covariant spin vector $\Vvec^\mu$ whose scalar product with itself
preserves Lorentz invariance for both inertial and noninertial motion, with interesting Lie algebra properties.
While there are potential experimental consequences which follow from the stated observations for $\Wvec^\mu$, the quantitative
relevance of the effects appears to be negligible for current experimental conditions.
There may be other ways of identifying the presence of $\Rvec$ at least on a theoretical level,
and it seems worth exploring how to make use of this formalism for other relevant physical situations,
such as particle interactions near the event horizon of a black hole.
At present, the theoretical results presented here suggest the existence of a new avenue for better understanding
the interface between quantum mechanics and curved space-time, via the principle of equivalence, when applied to
particles in noninertial motion.

In this paper, the consequences of noninertial motion on covariant quantum spin were explored by performing explicit calculations for
spin-1/2 particles.
For future considerations, it is worth mentioning that the noninertial effects of the type discussed in the present work
are also calculable for particles of higher intrinsic spin by employing appropriate representations of the Pauli-Lubanski vector \cite{Kirchbach}.
Other possibilities may come about from future study of these ideas.


\section{Acknowledgements}

The authors wish to thank Bahram Mashhoon for his careful review of the manuscript and helpful comments to
improve its overall quality.
We also thank Bhaskar Dutta and Saurya Das for useful discussions.
Funding for this research was provided in part by the Natural Sciences and Engineering Research Council (NSERC) of Canada.

\end{document}